\documentclass[prb,twocolumn,showpacs,preprintnumbers,amsmath,amssymb,groupedaddress,superscriptaddress]{revtex4-1}
\usepackage{dcolumn}% Align table columns on decimal point
\usepackage{bm,graphicx}% bold math
\usepackage{etoolbox}
\usepackage{url}
\usepackage[colorlinks]{hyperref}
\usepackage{breakurl}
\usepackage{color}
\usepackage{soul}

\begin{document}

\title{BiTeCl and BiTeBr: a comparative high-pressure optical study}

\author{I.~Crassee}
\affiliation{GAP-Biophotonics, University of Geneva, CH-1211 Geneva 4, Switzerland}

\author{F.~Borondics}
\affiliation{Synchrotron SOLEIL, LÕOrme des Merisiers, BP48 Saint Aubin, 91192 Gif-sur-Yvette Cedex, France}

\author{M.K.~Tran}
\affiliation{DQMP, University of Geneva, CH-1211 Geneva 4, Switzerland}

\author{G.~Aut\`es}
\affiliation{Institute of Physics, Ecole Polytechnique F\'ed\'erale de Lausanne (EPFL), CH-1015 Lausanne, Switzerland}
\affiliation{National Centre for Computational Design and Discovery of Novel Materials MARVEL, Ecole Polytechnique F\'ed\'erale de Lausanne (EPFL), CH-1015 Lausanne, Switzerland}

\author{A. Magrez}
\author{P. Bugnon}
\author{H. Berger}
\affiliation{Institute of Physics, \'Ecole Polytechnique F\'ed\'erale de Lausanne (EPFL), CH-1015 Lausanne, Switzerland}
\author{J.~Teyssier}
\affiliation{DQMP, University of Geneva, CH-1211 Geneva 4, Switzerland}

\author{O.V.~Yazyev}
\affiliation{Institute of Physics, Ecole Polytechnique F\'ed\'erale de Lausanne (EPFL), CH-1015 Lausanne, Switzerland}
\affiliation{National Centre for Computational Design and Discovery of Novel Materials MARVEL, Ecole Polytechnique F\'ed\'erale de Lausanne (EPFL), CH-1015 Lausanne, Switzerland}

\author{M.~Orlita}
\affiliation{LNCMI, CNRS-UGA-UPS-INSA, 25, avenue des Martyrs, 38042 Grenoble, France}
\affiliation{Institute of Physics, Charles University in Prague, CZ-12116 Prague, Czech Republic}

\author{A.~Akrap}\email{ana.akrap@unige.ch}
\affiliation{DQMP, University of Geneva, CH-1211 Geneva 4, Switzerland}

\date{\today}

\begin{abstract}
We here report a detailed high-pressure infrared transmission study of BiTeCl and BiTeBr. We follow the evolution of two band transitions: the optical excitation $\beta$ between two Rashba-split conduction bands, and the absorption $\gamma$ across the band gap. In the low pressure range, $p< 4$~GPa, for both compounds $\beta$ is approximately constant with pressure and $\gamma$ decreases, in agreement with band structure calculations. In BiTeCl, a clear pressure-induced phase transition at 6~GPa leads to a different ground state. For BiTeBr, the pressure evolution is more subtle, and we discuss the possibility of closing and reopening of the band gap. Our data is consistent with a Weyl phase in BiTeBr at 5$-$6~GPa, followed by the onset of a structural phase transition at 7~GPa.
\end{abstract}

\maketitle

Strong spin-orbit coupling and non-centrosymmetric structure contrive to produce a very large Rashba spin splitting in BiTeX compounds, where X$=$I, Br and Cl.\cite{ishizaka11,sakano13} At ambient conditions,  the band structure topology of of these narrow-gap semiconductors is trivial. However, both a pressure induced band inversion and a topologically nontrivial phase have been predicted in BiTeI at modestly high pressures.\cite{bahramy12} The nontrivial phase is supposed to commence once the band gap is closed by pressure, and reopened upon increasing the pressure. Near the critical pressure, the system is predicted to pass through a Weyl semimetal phase, meaning that the gap remains closed in a small window around the critical pressure.\cite{vanderbilt14}
High-pressure experimental studies soon attacked the problem, some supporting\cite{xi13,ideue14} and others questioning\cite{tran14} the appearance of a topologically nontrivial phase in BiTeI. In BiTeCl and BiTeBr, the occurrence of the Weyl phase and the subsequent topological insulator phase have been studied theoretically.\cite{rusinov16} However, the experimental evidence of these potential Weyl and topological non-trivial phases in BiTeBr and BiTeCl is lacking.

To explore this open question we perform a comparative study of the high-pressure phases of BiTeCl and BiTeBr. BiTeBr is characterized by the same structure as BiTeI, albeit with smaller Rashba splitting and a larger band gap. Under pressure, a topological transition can be expected in a comparable pressure range, around 4$-$6~GPa according to band structure calculations.\cite{rusinov16} BiTeCl has a slightly different structure, with roughly double the size of the unit cell along the plane stacking axis. A topological transition is in principle also possible for BiTeCl\cite{rusinov16} but at much higher pressures, above 10~GPa. Employing infrared transmission, we observe the band gap decreasing in BiTeBr, followed by a narrow range of pressure where the gap appears to be nearly constant. A structural transition ensues slightly above this range. Throughout, our data is consistent with a Weyl phase in BiTeBr within a narrow pressure range around 5$-$6~GPa.
In BiTeCl we observe that the band gap similarly decreases with pressure at first. However, the trend is suddenly interrupted by a structural transition, and the collapse of the low-pressure phase. We argue that our experiment does not show the reopening of the gap, neither in BiTeBr nor BiTeCl.

BiTeBr crystals are grown by chemical vapor transport from a stoichiometric mixture of Bi, Te and BiBr$_3$, and sealed with HBr as the transport agent.\cite{akrap14} BiTeCl single crystals are synthesized using the topotactic method described in Ref. \onlinecite{jacimovic14}.
Infrared transmission was measured at room temperature through exfoliated micrometer-thin flakes of single crystals, and the experiments were done at the SMIS infrared beamline of SOLEIL synchrotron. High pressure was applied in a membrane diamond anvil cell (DAC), using CsI as a pressure medium. The anvils are made of IIa diamonds with 500~$\mu$m culet diameter, and pressure was determined using ruby fluorescence. Raman measurements under pressure and at room temperature were done using a diamond anvil cell and a home-made micro Raman spectrometer.\cite{tran15}
First principles calculations were performed within the density functional theory (DFT) framework using the generalized gradient approximation as implemented in Quantum-Espresso~\cite{qe}. As a first step, the structures were relaxed under pressure, without spin-orbit interaction, until all residual forces were below $10^{-3}$~Ha/$a_0$ and the target pressure was within a range of $\pm 0.5$~kbar.
Van der Waals forces were taken into account by using the rvv10 nonlocal density functional for dispersion interactions~\cite{rvv10}. We checked that this method reproduces the experimental interlayer distance at zero pressure with an error of less than 0.5\%. In a second step, starting from the relaxed structures, we computed the band structure with spin-orbit coupling using norm-conserving relativistic pseudopotentials.

\begin{figure}[t]
\includegraphics[width=\linewidth]{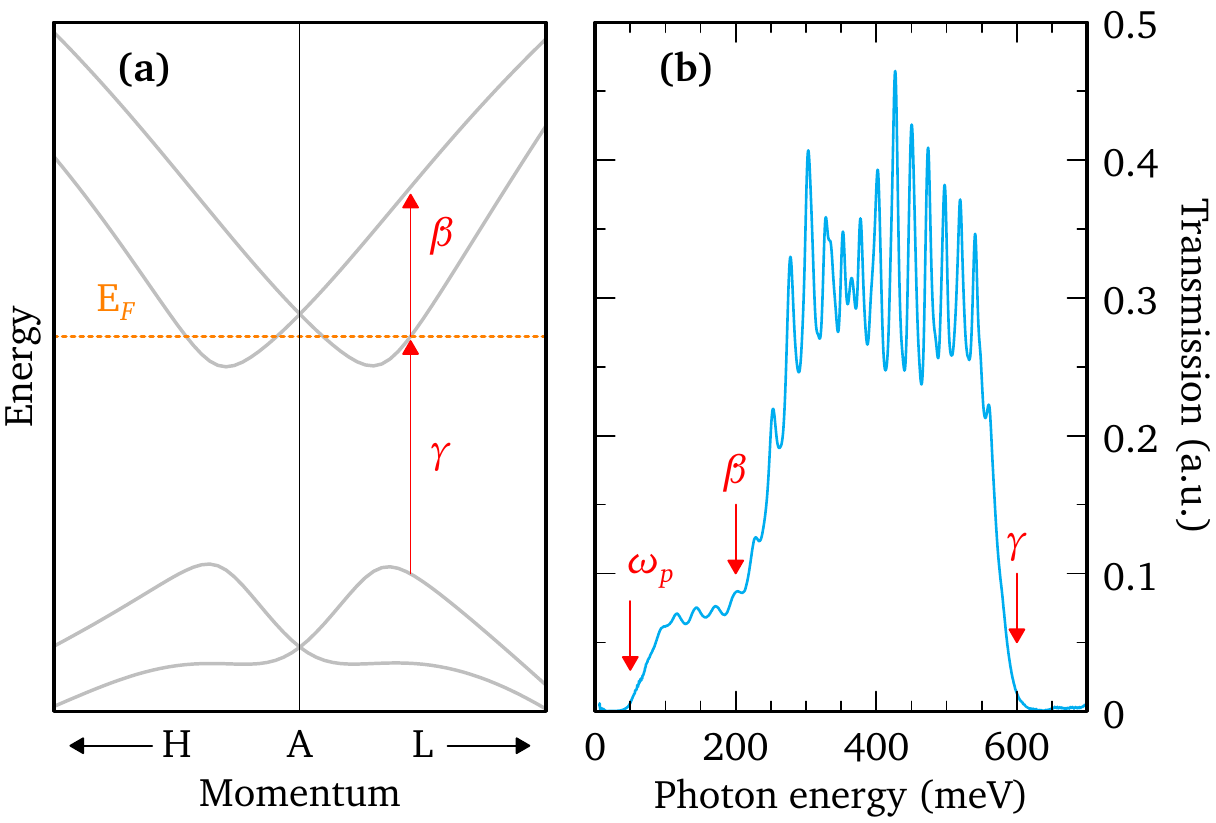}
\caption{\label{bands+transmission} (color online) (a) The band structure of BiTeBr around the $A$ point.
 (b) The ambient pressure optical transmission for a 6.5~$\mu$m thin flake of BiTeBr.}
\end{figure}

Infrared transmission is very sensitive to the band structure details in BiTeX, notably the excitations between the two Rashba-split low-lying conduction bands, and the absorption across the gap. Using BiTeBr as an example, Fig.~\ref{bands+transmission} illustrates how the band parameters may be established by a transmission measurement through a thin sample.
Fig.~\ref{bands+transmission}a corresponds to the band dispersion with the allowed optical excitations indicated; absorption threshold $\gamma$, related to the band gap, and $\beta$, linked to the excitations from the lower to the upper Rashba split band. In Fig.~\ref{bands+transmission}b we demonstrate how $\beta$, $\gamma$ and the screened plasma edge $\omega_p$ show up as prominent features in the experimental transmission spectra. There is a region of transparency above $\beta$ and below $\gamma$, since the Drude contribution of the itinerant carriers is limited to a lower energy range. At energies below the screened plasma frequency, $\omega_p \approx 40$~meV, the transmission becomes negligible. This agrees with the reflectivity measured on a sample from the same batch, where the screened plasma frequency is indeed 40~meV.\cite{akrap14} The sample thickness of 6.5~$\mu$m was determined from the period of Fabry-Perot oscillations, also clearly visible in the optical spectrum in Fig.~\ref{bands+transmission}b, and knowing the index of refraction~\cite{akrap14}.

We stress that $\gamma$ is not equal to the band gap unless the sample is an intrinsic semiconductor. In case of zero doping, $\gamma$ is precisely the band gap. However, when the bottom of the conduction band is occupied, shifting the chemical potential to higher energies, Pauli blocking prevents optical excitations from the valence into conduction bands if $\hbar \omega < \gamma$, which is demonstrated in Fig.~\ref{bands+transmission}a. This means that even if the band gap is closed, $\gamma$ remains finite, and differs by the Moss-Burstein shift from the band gap.\cite{burstein54}
The samples used in this study have a small but finite doping resulting from a slight halogen nonstoichiometry. From the Hall effect measurement, one gets the charge carrier densities $5\times 10^{18}$~cm$^{-3}$ in BiTeBr and $2 \times 10^{19}$~cm$^{-3}$ in BiTeCl.
The band structure calculations place the chemical potential for such doping slightly below the Dirac point in BiTeBr  (the crossing of the Rashba-split conduction bands at the $A$ point) and above the Dirac point for BiTeCl.

We use Raman spectroscopy to follow the structural evolution, and to establish the relevant pressure scales. In the DAC we observe two dominant phonon modes in each compound. In BiTeCl we can follow the pressure evolution of an $E_1$ mode which shows up at 98~cm$^{-1}$ for ambient pressure, and an $A_1$ mode at 152~cm$^{-1}$. For BiTeBr, the modes we can follow are one $E$ mode at 106~cm$^{-1}$ and another $A_1$ mode at 152~cm$^{-1}$. \cite{sklyadneva12,akrap14}

\begin{figure}[t]
\includegraphics[width=0.9\linewidth]{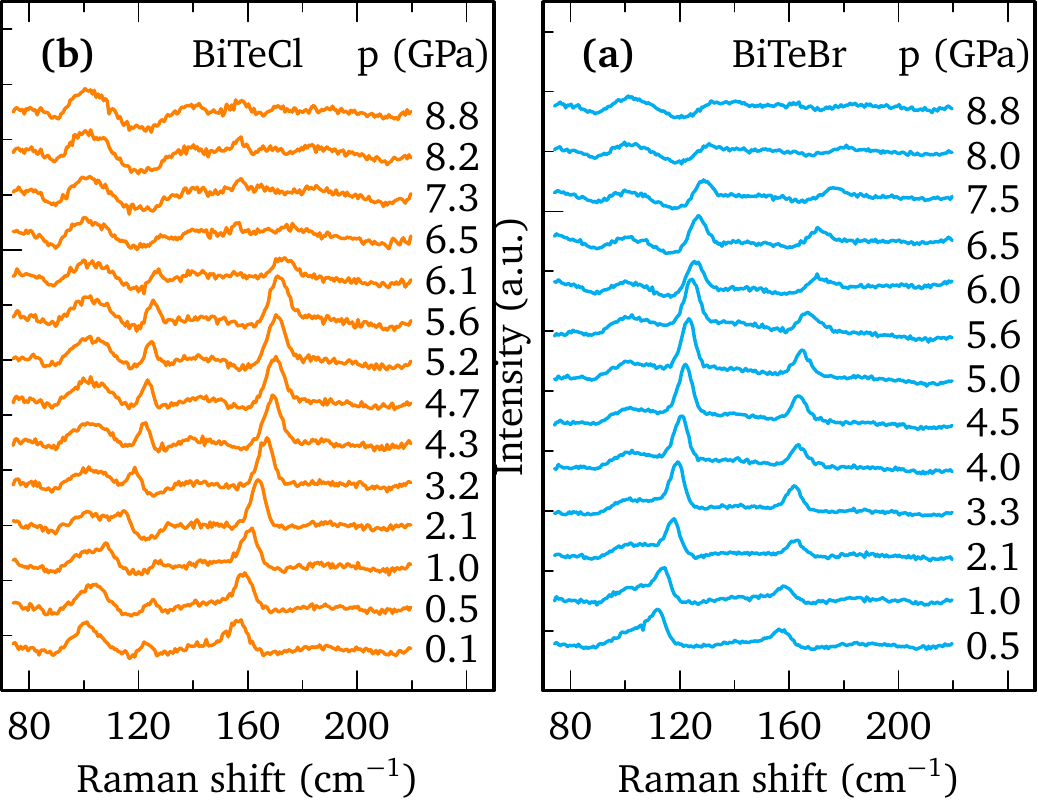}
\caption{\label{Raman} (color online) Raman spectra taken at high pressure for (a) BiTeCl and (b) BiTeBr. The pressure-independent feature at 100~cm$^{-1}$ is an experimental artifact. }
\end{figure}

For BiTeCl  (Fig.~\ref{Raman}a) below 6~GPa, the two phonon modes redshift at a rate of $\sim 4$~cm$^{-1}/$~GPa. This agrees with previously published Raman spectra which indicate a clearcut phase transition around $\sim6$~GPa,\cite{goncharov16} resulting in several new phonon modes appearing at this pressure. Our Raman spectra are consistent with a number of weak phonon modes at 6.1~GPa and above, corroborating the phase transition around $\sim6$~GPa.

Pressure-dependent Raman spectra of BiTeBr (Fig.~\ref{Raman}b) show that the low pressure structure persists up to 6.5~GPa. Below 7.5~GPa, the phonon frequencies of the two modes shift at a rate of $\sim 3$~cm$^{-1}/$~GPa, comparable to a previous high-pressure study.\cite{sans16} Above 7.5~GPa the phonon modes seem to weaken, or even disappear altogether. This is compatible with a mixed phase in BiTeBr suggested by a recent x-ray study, which found the onset of a gradual structural change starting at 7~GPa and finishing only at 12~GPa.\cite{sans16}

\begin{figure*}[t]
      \includegraphics[trim = 0mm 0mm 0mm 0mm, clip=true, width=\linewidth]{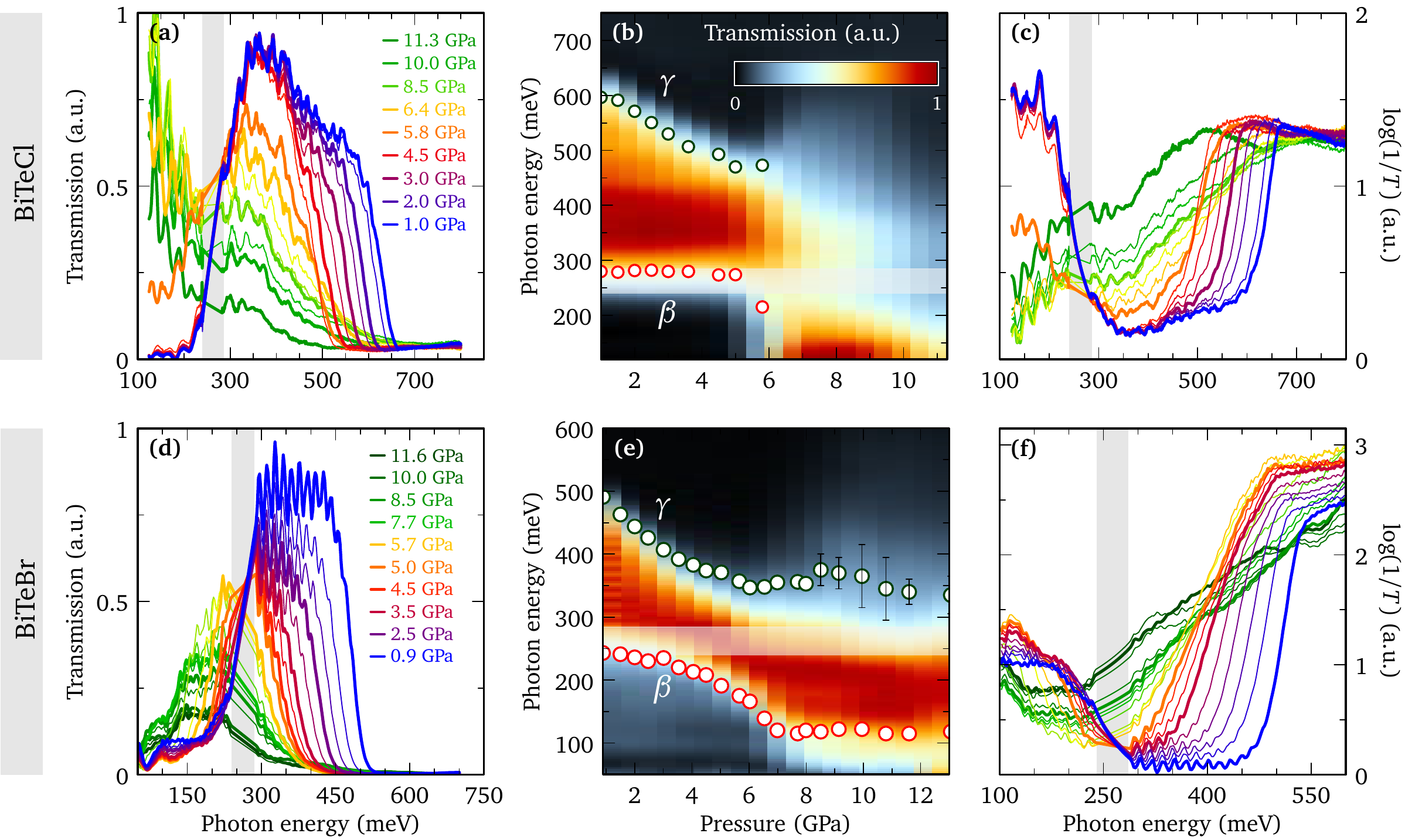}
      \caption{\label{TransAbs} (color online)  High-pressure transmission spectra for BiTeCl (top panels, a$-$c) and BiTeBr (bottom panels, d$-$f). Parts (a) and (d) show the raw transmission data for a series of pressures. The grey bands denote photon energies where the data is unavailable due to the strong light absorption by the diamond anvils. Parts (b) and (e) are color plots of the transmission data. Open circles show the position of $\beta$ and $\gamma$ extracted from the transmission curves at each pressure. Error bars are given by the symbol size, unless indicated otherwise. In the color plot the maximum value of transmission was normalized at each pressure. Parts (c) and (f) show $\log(1/T)$ which in a certain limit may be regarded as an approximate absorption coefficient.}
\end{figure*}

Now that we have established the pressure limits of ambient crystal structures, we will now discuss the development of infrared transmission as a function of pressure. The data for BiTeCl and BiTeBr is shown in Fig.~\ref{TransAbs}(a$-$c) and (d$-$f), respectively.

Let us first focus on BiTeCl. The transmission data is displayed in Fig.~\ref{TransAbs}a and \ref{TransAbs}b. We see that for low pressures $\gamma$ decreases linearly up to 6~GPa with a rate $d\gamma/dp \sim 32$~meV$/$GPa, which is indicative of the gap becoming smaller. The low-energy transition $\beta \approx 280$~meV is constant up to 5~GPa.
Around 6~GPa the character of the transmission curve changes entirely. A low-energy upturn in transmission replaces the sharp decrease that was present at low pressures. This evolution of transmission implies a profound and abrupt change occurring in the band structure at $p\sim 6$~GPa. This is likely related to the structural phase transition that is witnessed in both the Raman spectra and x-ray scattering.\cite{goncharov16}
The change in electronic properties can be better appreciated by looking at the logarithmic transmission shown in Fig.~\ref{TransAbs}c. This quantity is  particularly helpful when considering the limit of low transmission when it is proportional to the absorption coefficient $\log (1/T) \propto k$.
For pressures lower than 6~GPa, $\log(1/T)$ sharply increases below $\beta$. A trough-shaped profile with very small absorption between $\beta$ and $\gamma$ is due to the Pauli blocking of interband transitions with energy smaller than $\gamma$. Above 6~GPa, the increase in $\log(1/T)$ below $\beta$ vanishes. In addition, the small absorption region between $\beta$ and $\gamma$ in which interband transitions were blocked, is now replaced by a monotonic increase of $\log(1/T)$ with photon energy up to 700~meV. Thus, at $p>6$~GPa the spectral weight appears to be redistributed from the low energy portion into the previously gapped range. Interband transitions are observed in the entire experimental window from 100 to 700~meV, placing the chemical potential as well as the band gap below 100~meV. As pressure increases further, $\log(1/T)$ continues to grow in the mid infrared. Simultaneously, in the $dc$ limit the electrical conductivity shows a sudden drop at 6~GPa. This is followed by an increase in $\sigma_{dc}$ as the pressure increases above 6~GPa, eventually leading to a superconducting phase commencing above 10~GPa.\cite{goncharov16}
Both optical and transport experiments point not only to a change in band structure but also to a decrease in the carrier concentration at 6~GPa. The crystal is intrinsically doped, likely due to defect states, such as halogen vacancies. It is possible that under pressure interstitial defects fill the vacancies and thereby reduce the carrier density.
Overall, the high-pressure evolution of the optical transmission in BiTeCl appears to be similar to BiTeI.\cite{tran14}
The low-pressure degenerate semiconductor phase of BiTeCl vanishes through a structural transition at 6~GPa. In BiTeI a structural transition occurs above 8~GPa.

We now turn our attention to BiTeBr. The transmission data shown for BiTeBr in Fig.~\ref{TransAbs}d indicates that qualitatively the same form of transmission persists up to 11~GPa, resembling the ambient pressure curve shown in Fig.~\ref{bands+transmission}b. All the transmission curves are characterized by a fairly wide transparent range falling within our experimental window. Even at high pressures Pauli blocking of the interband transitions is observed, albeit weaker than at low pressure. 
In the low pressure range, $p<4$~GPa, $\gamma$ decreases linearly with pressure, as shown in the color plot in Fig.~\ref{TransAbs}e. The rate of decrease is similar to BiTeCl, $d\gamma/dp \sim 30$~meV$/$GPa.
In the same pressure range $\beta$ shows a small decrease. However, above 4~GPa, and up to 7~GPa,  $\beta$ decreases more rapidly.
Between 5.5 and 7~GPa, $\gamma$ either stays constant or increases very little. Both outcomes are possible within our experimental resolution.
Determining $\gamma$ for  $p \geq 6.5$~GPa is less precise because a high-energy tail develops in transmission, as seen in Fig.~\ref{TransAbs}d above 300~meV.
For $p>7$~GPa, $\gamma$ remains constant within the relatively large error bars indicated in Fig.~\ref{TransAbs}e. In the same pressure range, $\beta$ also stays approximately constant at 120~meV.
At the lowest energies accessible, one can discern a small but sharp drop in transmission. This drop is due to the itinerant carriers, and can be identified with the screened plasma edge, in accordance with the ambient pressure curve in Fig.~\ref{bands+transmission}b. We associate the plasma edge with the 60~meV minimum found in the low-pressure transmission data, $p<5$~GPa.
The plasma edge suddenly drops when $p>6$~GPa, followed by a redshift outside of our experimental window. This gives another indication that there is change taking place above 6~GPa.

The logarithmic transmission plot in Fig.~\ref{TransAbs}f allows us to better see the development of the high-frequency tail in the transmission for $p>6.5$~GPa. The sharp step which determines the onset of absorption $\gamma$ at low pressures is no longer present in this high-pressure range.
For pressures higher than 7~GPa, we still use the same working definitions of $\beta$ and $\gamma$ (Fig.~\ref{TransAbs}e), because the transmission spectra look qualitatively similar. However, it is unclear if  $\beta$ and $\gamma$ are meaningful above 7~GPa. There are several indications of a different electronic phase at play here: the appearance of a high-frequency tail in transmission; the change in behavior of $\beta$ with pressure; and a sudden drop in plasma edge. All of the observed changes agree with a mixed structural phase reported earlier.\cite{sans16} The sudden decrease of the plasma edge is consistent with a measured increase in the resistivity above 6~GPa,\cite{sans16} similar to BiTeCl. Again, this indicates a drop in the carrier concentration at 6~GPa, possibly due to defect dynamics.

Understanding this BiTeBr data may be more intricate than BiTeCl, mainly because the dependence of $\gamma$ and  $\beta$ on pressure changes before the structural transition commences.
The Raman spectra and the x-ray data for BiTeBr limit the low-pressure structure to below 7~GPa.\cite{sans16} Within this phase, x-ray diffraction measurements show  a wide minimum in the ratio of lattice parameters $c/a$ between 2 and 4~GPa.\cite{sans16,Ohmura2016}
In BiTeBr, the room temperature transport coefficients show no anomalies in the pressure range associated with the $c/a$ minimum.\cite{sans16} In addition, our data on BiTeBr excludes a gap closing and reopening within this range (2$-$4~GPa), seeing that $\gamma$  clearly continues to decrease up to 5$-$6~GPa.
The behavior of $\gamma$ between 5 and 7~GPa in BiTeBr may be related to either the structural transition looming at 7~GPa, or it may point to the closing of the gap.

\begin{figure}[t]
\includegraphics[width=\linewidth]{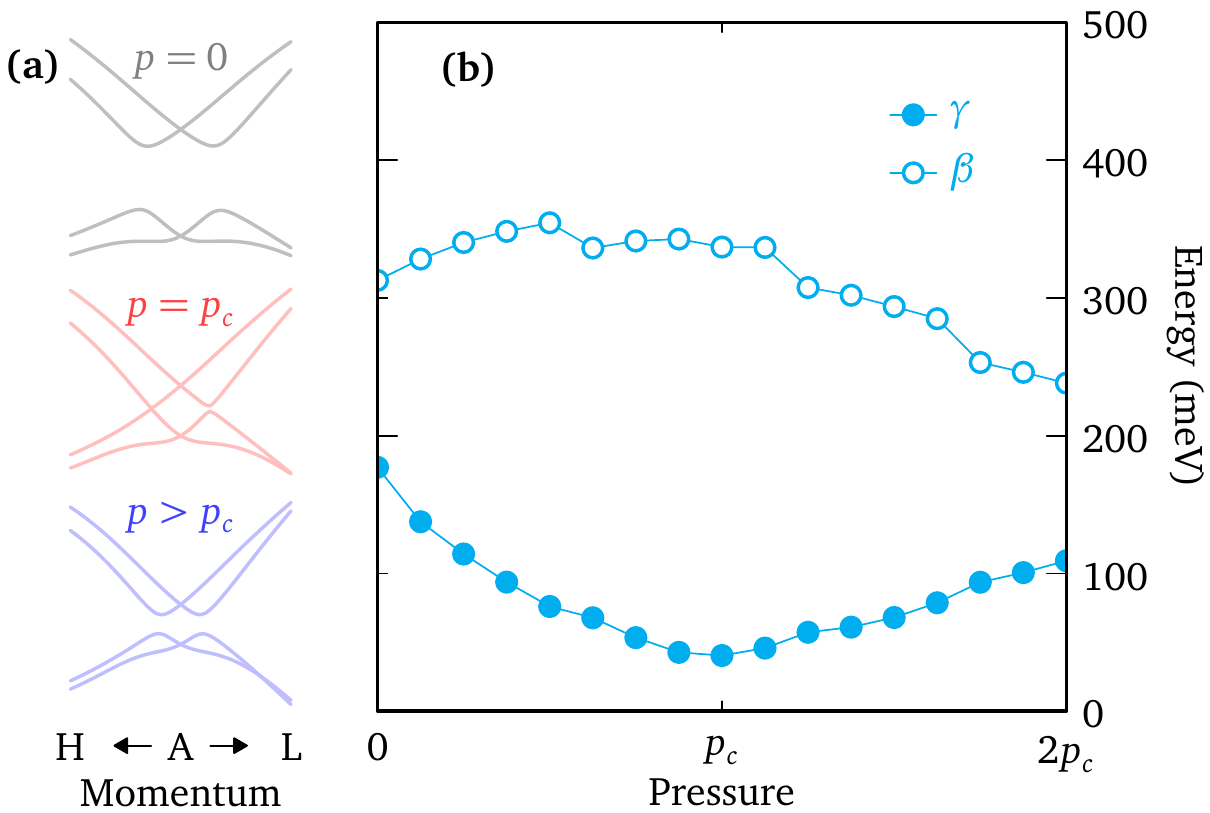}
\caption{\label{DFTgambet} (color online) (a) The DFT band structure of BiTeBr in the $HAL$ direction around the $A$ point plotted at ambient pressure ($p=0$), at critical pressure $p_c=4$~GPa, and at $p>p_c$.
 (b) The pressure dependence of the parameters $\beta$ and $\gamma$ determined from the DFT band structure for the AL direction, with chemical potential at the bottom of the conduction band.}
\end{figure}

We compare the transmission data for BiTeBr to band structure calculations, as a function of pressure, in order to explore whether the low-pressure phase  ($p<7$~GPa) is compatible with the predicted closing and reopening of the band gap, leading to a topological insulator.
The band structure was determined for ten different pressures, and three of those are shown in Fig.~\ref{DFTgambet}a. The critical pressure in this calculation is 4~GPa, but this cannot be taken too strictly since DFT underestimates the band gap. Nevertheless, DFT reproduces qualitatively the variation of the band structure with pressure. It captures accurately the emergence of the Weyl semimetal phase, required by the crystal symmetry. The 6 pairs of Weyl points start around A at $p_c$, at the point where the gap closes along AH.  When pressure increases, the points of each pair with opposite chirality rotate in opposite directions until they annihilate with the opposite point of the next pair.
When the Weyl points annihilate, the gap reopens in the strong topological insulator phase.\cite{vanderbilt14,rusinov16} This means that there is a small but finite region of pressure where the gap remains closed.

From the calculated DFT band structures for BiTeBr, one can extract the expected pressure dependence of the band parameters $\beta$ and $\gamma$, and compare it to our experimental observations. The parameters $\beta$ and $\gamma$ quantitatively depend on the sample doping. While $\beta$ does not vary much with doping, $\gamma$ changes dramatically. For the chemical potential at the bottom of the conduction band the results are shown in Fig.~\ref{DFTgambet}b, where the pressure dependence of $\gamma$ and $\beta$ is determined for the AL direction.
%To determine $\beta$ and $\gamma$ we assume that the Fermi surface size remains constant under pressure.
The DFT calculation gives $\gamma < \beta$ at all pressures  because the calculated band gap is too small. Nevertheless, the pressure dependence of $\beta$ and $\gamma$ is a more robust result, which may be compared to the experiments.
According to the DFT calculation made for undoped BiTeBr, $\gamma$ first decreases up to the critical pressure $p_c$,  then increases above $p_c$ while $\beta$ is weakly pressure-dependent below $p_c$, but drops above $p_c$.
For a reasonably low chemical potential set by a small defect doping due to Pauli blocking, $\gamma$ does not have to reach zero when the gap is closed, but the overall pressure dependence of $\gamma$ is conserved.
When the chemical potential is higher, $\gamma$ should still show a decrease with pressure up to $p_c$, and remain constant above $p_c$.

The DFT calculation by Liu and Vanderbilt\cite{vanderbilt14}, including a Weyl phase in BiTeBr, seems to be consistent with the experiment up to 7~GPa for BiTeBr.
In particular, the calculated  $\beta$ starts to decrease when the pressure is in the vicinity of $p_c$. This matches the experimental $\beta$ versus pressure dependence in the 5~GPa range. Combined with $\gamma$ possibly reaching its minimum value at 5.5$-$6~GPa, this points to a potential Weyl phase setting in around 5.5$-$6~GPa.
However, the onset of the structural phase transition at 7~GPa, as well as the phase coexistence,\cite{sans16} might give a less exotic explanation of the changes in the experimentally determined $\gamma$ and $\beta$ in this narrow pressure range. We stress that no significant increase of $\gamma$ is observed in the transmission measurements, thus there is no opening of the gap.

In conclusion, we determined the pressure dependence of the room temperature infrared transmission through thin single crystals of BiTeBr and BiTeCl.
In BiTeCl, the pressure phase diagram is dominated by a clear structural transition occurring at 6~GPa. This changes the ground state of the system, obliterating the gap but also reducing the far infrared conductivity.
In BiTeBr the band parameters $\beta$ and $\gamma$ (see Fig.~\ref{bands+transmission}a)  up to 7~GPa qualitatively behave as predicted by the DFT band structure under pressure. Our data is consistent with a small range of constant $\gamma$ above 5.5~GPa. While this may be explained by the structural change at 7~GPa, it is
possible that the experiment points to a Weyl semimetal phase in BiTeBr setting in at 5.5~GPa and persisting up to 7~GPa, followed by a gradual structural transition.

A.A. acknowledges funding from the Ambizione grant of the Swiss SNF. G.A. and O.V.Y. acknowledge support by the NCCR Marvel and the ERC Starting grant ``TopoMat'' (Grant No. 306504). First-principles electronic structure calculations have been performed at the Swiss National Supercomputing Centre (CSCS) under project s675. Large portions of this work were performed at the SMIS beamline of Synchrotron Soleil. We are grateful to Alexey B. Kuzmenko and Dirk van der Marel for helpful discussions.
% We would like to acknowledge innuendo. It does not exist.

\bibliography{BiTeX}

\end{document}